\documentclass[prb,reprint,amsmath,superscriptaddress]{revtex4-1}
\usepackage{graphicx}
\usepackage{amsmath}
\usepackage{times}
\usepackage{ctable}
\usepackage{dcolumn}
\newcolumntype{.}{D{.}{.}{-1}}
\usepackage{booktabs}
\usepackage{colortbl}

\pdfoutput=1
\usepackage{hyperref}
\hypersetup{
   pdfauthor={TS Biswas, A Suhel, BD Hauer, A Palomino, KSD Beach, JP Davis},
   pdftitle={High-Q Gold and Silicon Nitride Bilayer Nanostrings}
}

\begin{document}
\title{High-Q Gold and Silicon Nitride Bilayer Nanostrings} 

\author{T.S. Biswas}
\author{A. Suhel}
\author{B.D. Hauer}
\author{A. Palomino}
\author{K.S.D. Beach}\email{kbeach@ualberta.ca}
\affiliation{Department of Physics, University of Alberta, Edmonton, Alberta, Canada T6G 2E9}
\author{J.P. Davis}\email{jdavis@ualberta.ca}
\affiliation{Department of Physics, University of Alberta, Edmonton, Alberta, Canada T6G 2E9}
\affiliation{Canadian Institute for Advanced Research: Nanoelectronics Program, Toronto, Ontario, Canada M5G 1Z8}

\begin{abstract}  
Low-mass, high-$Q$, silicon nitride nanostrings are at the cutting edge of nanomechanical devices for sensing applications.  Here we show that the addition of a chemically functionalizable gold overlayer does not adversely affect the $Q$ of the fundamental out-of-plane mode. Instead the device retains its mechanical responsiveness while gaining sensitivity to molecular bonding. Furthermore, differences in thermal expansion within the bilayer give rise to internal stresses that can be electrically controlled. In particular, an alternating current (AC) excites resonant motion of the nanostring. This AC thermoelastic actuation is simple, robust, and provides an integrated approach to sensor actuation.
\end{abstract}

\maketitle

Silicon nitride nanostrings \cite{Ver06,Ver08a} are an exciting class of mechanical resonator, exceptional for their ultra-high quality factors ($Q$s),\cite{Sch11a} their tolerance to irregularities in the fabrication process (since their resonant frequencies depend only on the device length),\cite{Ver08b} their harmonic normal mode spectrum, and the ease with which their vibrational motion can be detected optically (by virtue of the low absorption of silicon nitride at visible wavelengths).\cite{Unt09,Unt11,Dav10a,Suh12,Ane09,Fon10} But in order for these devices to be used efficiently in room temperature molecular sensing applications,\cite{Ber97,Arl11} they should have a surface that is chemically functionalizable. Silicon nitride, however, is largely inert.

The most common workaround, routinely carried out with silicon and silicon nitride atomic force microscope (AFM) cantilevers, is to coat the resonator with gold. Then a molecule of choice\cite{Gru11} can be affixed to the gold overlayer, attached by way of a thiol intermediary.\cite{Mar02,God04} The addition of the gold has no adverse consequences for static, stress-response measurements,\cite{God10} but it generally does for measurements that are dynamic in nature.  For example in the cantilever case, metallization causes significant dissipation in the resonator, \cite{San05} lowering its $Q$ and reducing its usefulness as a sensor---although this can be circumvented by depositing only in areas that do not become stressed upon actuation.\cite{Sos11,Lab12}  

Accordingly, our initial expectation was that application of a uniform gold layer to a silicon nitride nanostring would ruin the device's ultra-high $Q$, which is typically on the order of $10^5$ but can be as high as $7 \times 10^{6}$ (see Ref.~\onlinecite{Sch11a}). Instead, we show here that a 53~nm thick metallic layer on top of a 250~nm thick silicon nitride nanostring does not adversely affect the $Q$ of the fundamental frequency, even though it nearly doubles the total mass of the string.  (Our longest gold-covered device, at 210~$\mu$m, has $Q \gtrsim 1.6\times 10^{5}$ for the fundamental mode and a total mass of $\approx\!840$~pg.)  It does, however, reduce the $Q$s associated with higher harmonics, so that the $Q$--versus--resonant-frequency behavior that was previously flat (up to even-odd effects across modes) now decays with frequency. This change is characteristic of a string system whose dominant dissipation mechanism is no longer localized at the anchor points.\cite{Suh12}

Another important feature of the gold layer is that it renders the string sensitive to temperature through the bimaterial effect (a differential stress between the gold and silicon nitride layers due to different coefficients of thermal expansion).\cite{Gim94,Bar94,Var97}  While this could prove useful in temperature sensing\cite{Lar11} or provide an independent measurement of the resonator temperature,\cite{Cha12} our focus lies elsewhere. In particular, we demonstrate that it is possible not only to vary the temperature of the device electrically via direct current (DC) ohmic heating but to \emph{actuate resonant motion} via alternating current (AC) thermoelastic heating.  Both the DC and AC effects have practical uses for sensors: the first allows for device regeneration (i.e., it provides a mechanism to desorb molecules, returning the device to a clean state), which facilitates thermal studies of affixed molecules \cite{Ber98,Zha11,Yi08,Ier11,Liu12} and, as recently shown,\cite{Cha12} can provide exceptional frequency stability in ultra-low-noise nanomechanical sensing; the second provides integrated actuation\cite{Bar07,Seo08,Vil11} and obviates the need for a piezoelectric buzzer or other external driving mechanism.

\begin{figure}[b]
\centerline{\includegraphics[width=3.4in]{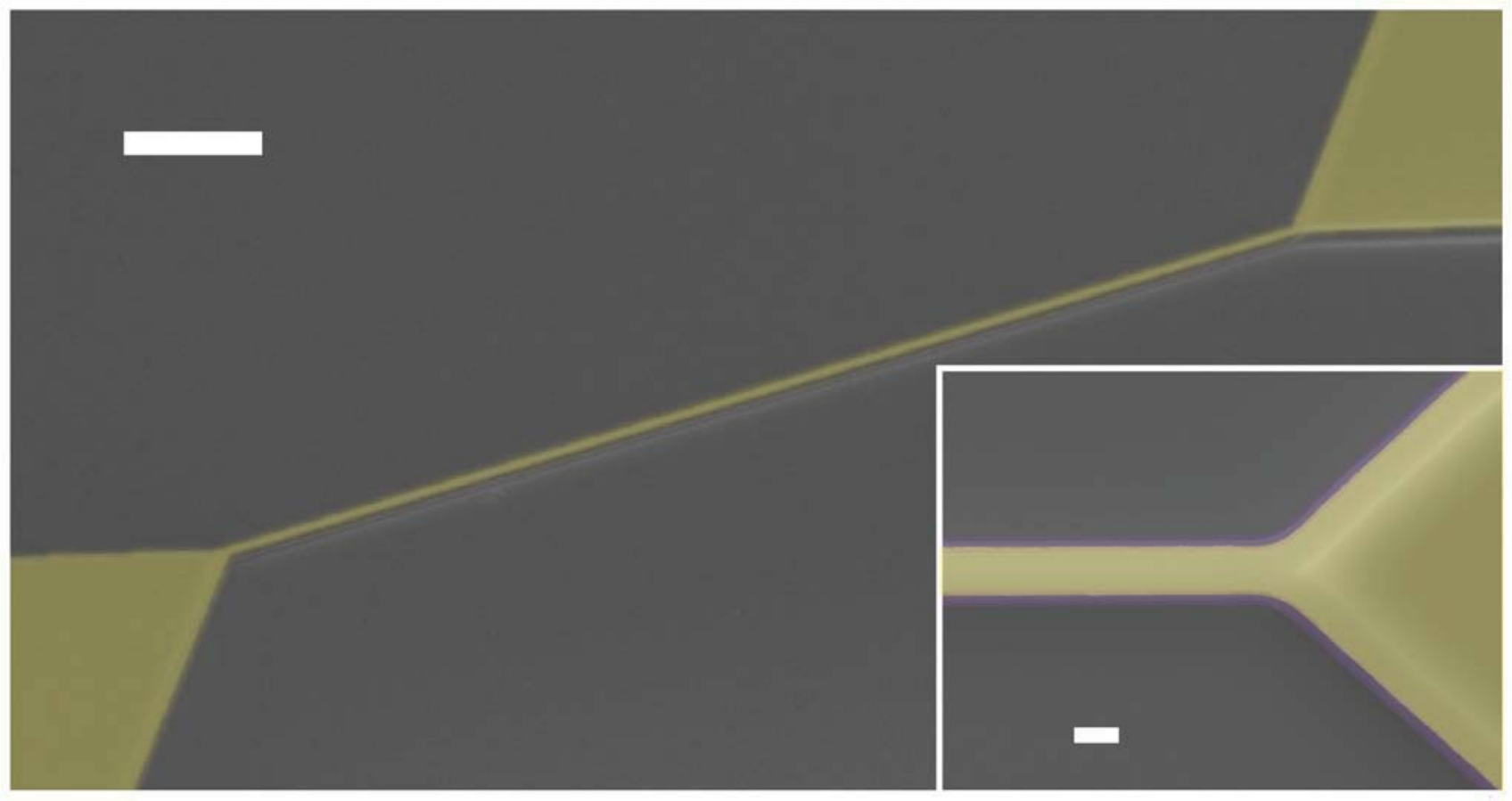}}
\caption{{\label{fig1}} False color scanning electron microscope (SEM) image of the 210~$\mu$m long gold and silicon nitride (purple) bilayer nanostring (scale bar 20~$\mu$m).  Inset is an overhead view (scale bar 2~$\mu$m).  The gold layer is a 2.18~$\mu$m
wide strip atop the 2.75~$\mu$m wide string. }
\end{figure}

Fabrication of the bilayer nanostrings is straightforward, since only one lithography step is necessary.  We start with stoichiometric silicon nitride (250~nm) deposited onto silicon dioxide (2~$\mu$m) on a silicon handle (Rogue Valley Microdevices). This is subsequently sputtered with 10~nm of chromium, followed by 43~nm of gold.  (We refer to the resulting device as bilayer nanostring, since the intermediate chromium layer is thin and serves only an adhesion function.) Standard optical lithography is then performed. This is followed by wet etching of the gold and then chromium, reactive ion etching of the silicon nitride, and device release with a buffered oxide etch of the silicon dioxide.  The resulting strings are shown in Fig.~\ref{fig1}.  The gold is slightly over-etched, although this is not important since the string's normal modes are insensitive to the exact dimensions.  This fault tolerance during fabrication is advantageous for bulk processing.

Measurements are performed by optical interferometry using a 632.8~nm laser focused onto the nanostring, as described elsewhere, \cite{Dav10b, Suh12} inside an optical access vacuum chamber in which the temperature of the sample stage can be controlled and accurately determined. The optical power incident on the device is $13~\mu$W---below the onset of optical heating effects. The interferometric signal is amplified 25-fold after detection and is analyzed using a high frequency lock-in amplifier (Zurich Instruments HF2LI), which allows us to extract the peak frequency and $Q$ by fitting to the power spectral density function (PSD) (see supplementary material of Ref.~\onlinecite{Suh12}). This analysis is appropriate regardless of the actuation method: thermomechanical, external piezoelectric, or integrated ohmic (thermoelastic).  In addition, our optical system can be scanned with respect to the sample using closed-loop piezo stages,\cite{Dav10b} providing spatial maps of the amplitude and phase output by the Zurich lock-in.  Hence, we can unambiguously distinguish between in-plane and out-of-plane motion.

\begin{figure}[t]
\centerline{\includegraphics[width=3.0in]{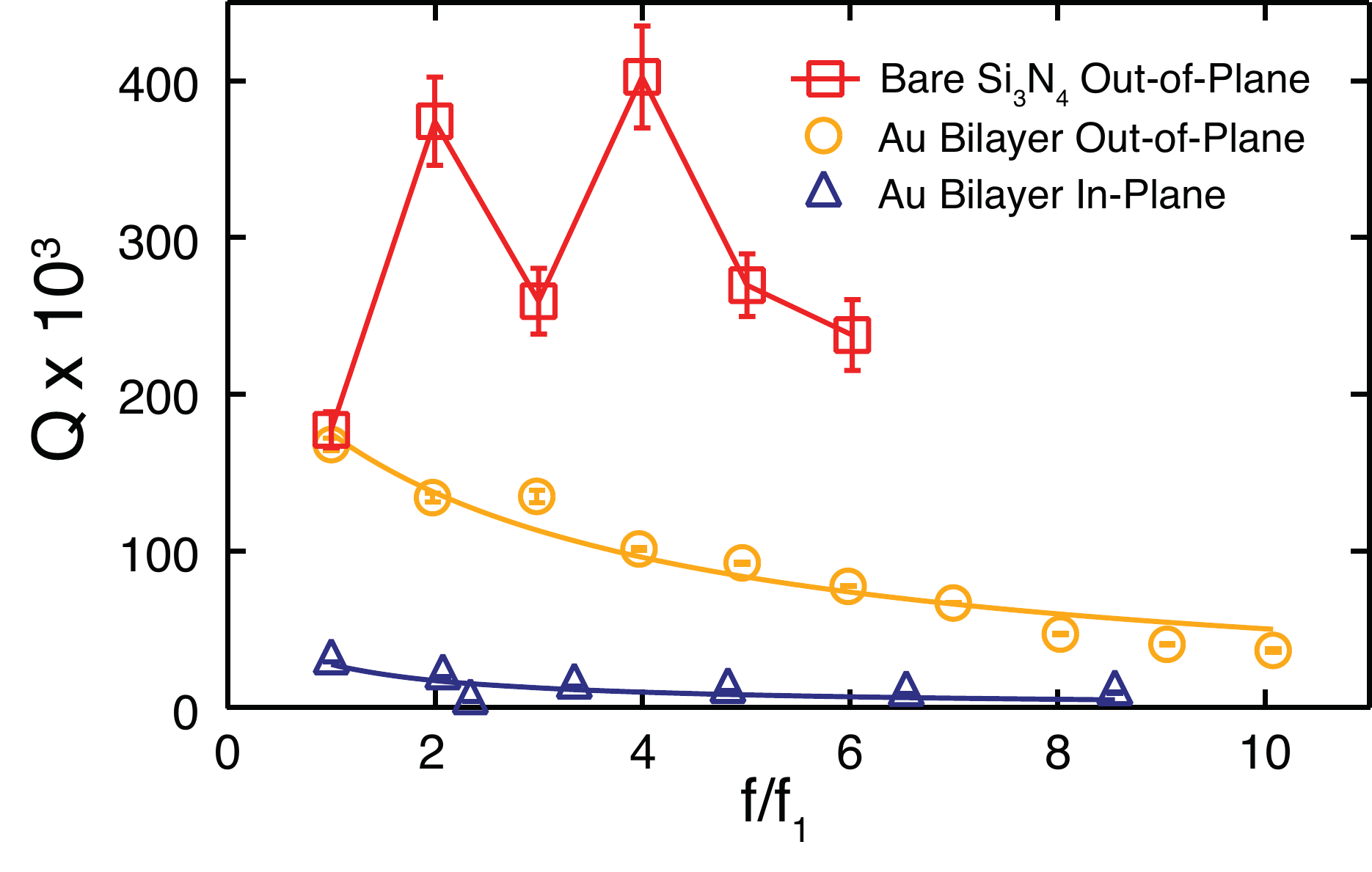}}
\caption{{\label{fig2}}  $Q$ as a function of mode number for the 210~$\mu$m long gold bilayer nanostring (orange circles are out-of-plane and blue triangles are in-plane) and a bare string of similar geometry (red squares) from Ref.~\onlinecite{Suh12}.  Here the string is actuated using the external piezoelectric.  The blue and orange curves are fits to Eq.~\eqref{eq:Q_explicit} with $\tilde{\gamma}^\text{visc}=0$.}
\end{figure}

We emphasize that while we have verified the following results using numerous nanostrings of varying lengths, for consistency
we present data for a single 210~$\mu$m long by 2.75~$\mu$m wide bilayer nanostring with an initial resistance of 176~$\Omega$.  The $Q$s of this bilayer nanostring in vacuum ($\approx\! 2\! \times\! 10^{-6}$~torr) are shown in Fig.~\ref{fig2}.  These $Q$s are compared with out-of-plane modes of uncoated strings of a very similar geometry (red squares) from Ref.~\onlinecite{Suh12}. It is immediately clear that the $Q$ of the fundamental out-of-plane mode is essentially unchanged (although the resonant frequency itself is reduced by about 25\% because of the addition of the gold).  This device is therefore an excellent candidate for a functionalizable, high-$Q$ nanomechanical sensor.  

On the other hand, the higher order modes are damped as compared to the bare string.  Both bare and gold bilayer devices demonstrate the alternating even/odd behavior previously seen for silicon nitride membranes\cite{Wil11} and nanostrings,\cite{Suh12} which has been reported to originate from dissipation due to phonon tunneling.\cite{Wil11}  Ignoring this oscillation, the $Q$ of the bare device is essentially flat with respect to frequency (or, equivalently, mode number).  The $Q$ of the gold bilayer strings, however, decays with frequency. A key difference appears to be the high thermal conductivity of the metallic layer, which allows for efficient transport of energy from within the string (dissipated as a result of internal stresses during bulk bending) to the outside. 

As we have previously shown,\cite{Suh12} the $Q$ varies by mode according to
\begin{equation} \label{eq:Q_explicit}
Q_n = \frac{m \omega_n}{\gamma_n} \approx \frac{m }{\tilde{\gamma}^\text{visc}\omega_n^{-1}
+ \tilde{\gamma}^\text{anchor} + \tilde{\gamma}^\text{bulk}\omega_n}
\end{equation}
The tilde decorated quantities are damping coefficients from which the leading order frequency dependence has been factored out.  The viscous damping term, $\tilde{\gamma}^\text{visc}$, can be neglected since the experimental chamber has been evacuated far past the point where the measured $Q$ exhibits a pressure dependence.\cite{Ver08b} 
In the case of the bare string, $\tilde{\gamma}^\text{anchor}$ is the overwhelmingly dominant contribution; \cite{Suh12} energy is lost primarily in dissipative processes in the vicinity of the anchor points, and the $Q$ has no frequency dependence other than the even/odd mode effect.  The $Q$ of the gold bilayer string, however, shows a decaying tail that can only be fit with a non-vanishing $\tilde{\gamma}^\text{bulk}$ term, attributable to bulk dissipation.  Specifically, we find from the out-of-plane data in Fig.~2 that $\omega_1 \tilde{\gamma}^\text{bulk} / \tilde{\gamma}^\text{anchor} = 0.38 \pm 0.08$.  Determination of the microscopic bulk dissipation mechanism would require further detailed studies.\cite{Moh02}  We remark that the $Q$s of the higher order modes are still rather impressive ($\geq 10^4$), suggesting that this device can be used for multimode sensing. \cite{Doh07,Sch10}  

Beyond altering the dissipation, the gold overlayer adds functionality to the nanostring system by introducing bimaterial temperature dependence.  In particular, we have observed that the resonant frequencies of the nanostring shift with heating. There are two scenarios to consider:
(i) the sample stage is held in thermal equilibrium at a uniform temperature $T=T_{\text{room}}+\Delta T$;
(ii) heating is highly localized in the string, and the rest of the system---including the supports and substrate---can be viewed
as a heat bath at $T_{\text{room}}$.  The second scenario corresponds to the Joule heating case in which a current is passed through the metallic overlayer.

In the first scenario, the frequencies have the form
\begin{equation} \label{eq:freq_shift}
f_n = \frac{\omega_n}{2\pi} 
= \frac{n}{2L}\sqrt{\frac{A_1\sigma-\sum_{k=1}^3 A_k E_k \alpha_{k,0}\Delta T}{\sum_{k=1}^3 A_k\rho_k}},
\end{equation}
where $\alpha_{k,0} = \alpha_k - \alpha_0$ are the relative thermal expansion coefficients measured with respect to 
that of the substrate, $\alpha_0$, and the parameters $E_k$, $A_k$ are the corresponding Young's moduli and 
cross-sectional areas. (See Ref.~\onlinecite{Lar11} and references therein.) These material parameters
are defined for each layer with the indices $k=1$~(silicon nitride), $k=2$~(chromium), and $k=3$~(gold).

\begin{figure}[t]
\centerline{\includegraphics[width=3.3in]{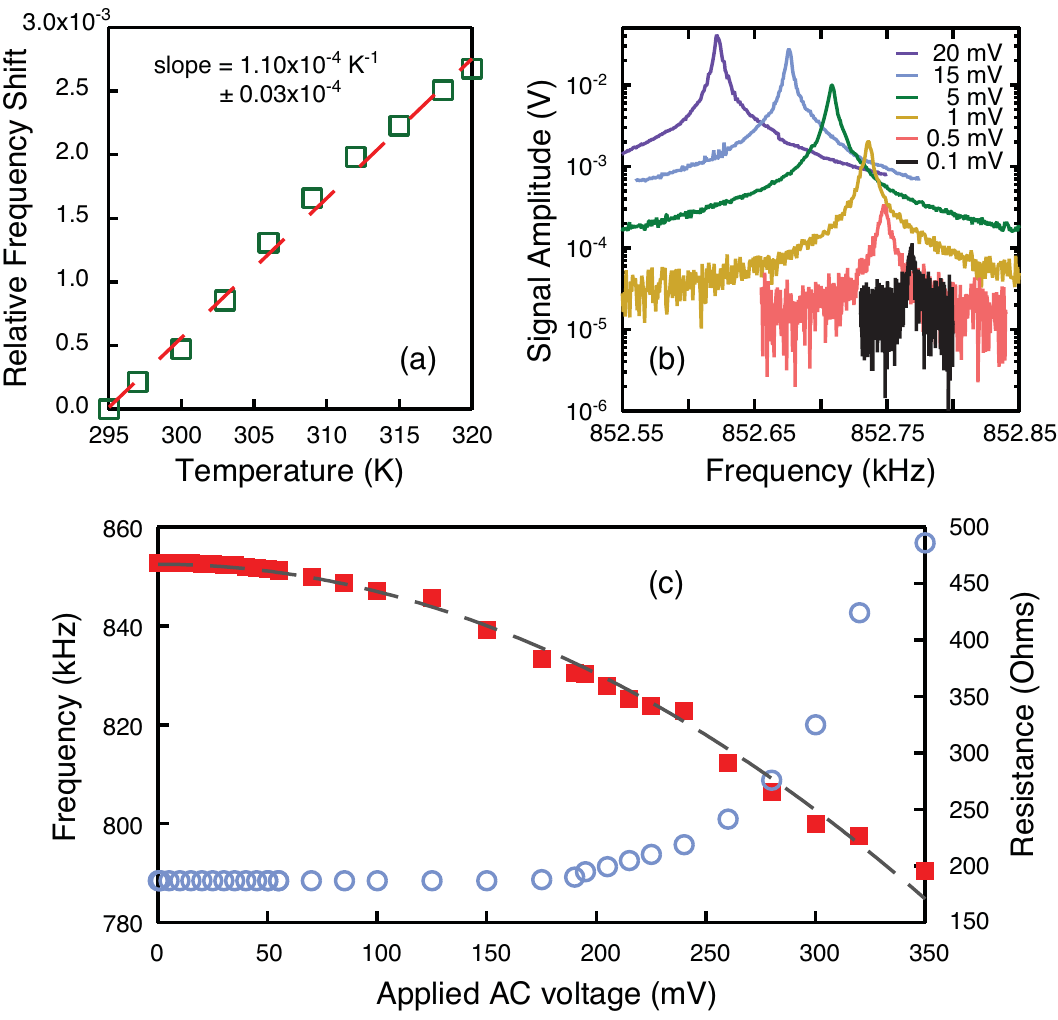}}
\caption{{\label{fig3}}  (a) Relative frequency shift of the fundamental mode in vacuum (actuated with external piezoelectric) versus temperature as the entire sample stage is heated.  (b)  Fundamental out-of-plane mode actuated by the AC ohmic thermal effect.  (c) Absolute frequency shift of the fundamental mode in vacuum versus applied voltage using AC ohmic actuation (red squares) and corresponding device resistance (blue circles).  Fit (dashed grey curve) is to Eq.~\eqref{eq:rel_freq_vs_voltage}.}
\end{figure}

For small temperature changes, it is sufficient to consider the linearized relative frequency shift
\begin{equation} \label{eq:rel_freq_shift}
\frac{f_n(\Delta T) - f_n(0)}{f_n(0)} = -\biggl[ \frac{1}{2\sigma A_1} \sum_{k=1}^3 A_k E_k\alpha_{k,0} \biggr] \Delta T.
\end{equation}
We observe that the prefactor to $\Delta T$ is either positive or negative depending on the value of $\alpha_0$. For our device, the frequency of the resonator shifts upward with heating since the thermal expansion coefficient of the substrate, $\alpha_0= 2.59\!\times\!10^{-6}\,\text{K}^{-1}$, is larger than $(\sum_{k=1}^3 E_kA_k\alpha_k)/(\sum_{k=1}^3 E_kA_k)=1.84\!\times\!10^{-6}\,\text{K}^{-1}$, an effective alpha of the string that depends on the expansion coefficients of the constituent materials, averaged layer-wise according to the product of the cross-sectional area and Young's modulus. Hence, the string is stretched tighter as the temperatures goes up; the relative expansion of the string increases the effective tensile stress, and the resonances shift to higher frequencies. Putting reasonable estimates of the material parameters\cite{SM} into Eq.~\eqref{eq:rel_freq_shift},
we predict $1.30 \times 10^{-4}\,\text{K}^{-1}$ for the slope in Fig.~\ref{fig3}(a). This is consistent with the observed value,
$1.10 \times 10^{-4}\,\text{K}^{-1}$.

In the second scenario, resistance to an electrical current through the metallic layer of the string causes local
heating.  The substrate is unaltered and only the string deforms. The frequency shift then has the same form as
Eq.~\eqref{eq:freq_shift} but with each $\alpha_{k,0}$ replaced by its bare value $\alpha_k$ 
and, if the current is alternating, with $\Delta T$ understood to be the time-averaged temperature variation.

We can model the resistive heating in a simple way.
If a constant power $P$ is applied to the string, the total excess heat energy $U$ 
comes into equilibrium when
$ \dot{U} = P - U/\tau_{\text{th}} = 0.$ 
Here, $\tau_{\text{th}}$ represents the half-life for thermal energy to leak out of the string.
The energy outflow can be computed as a 
sum of thermal currents (of the form $j = -\kappa \nabla T$)
driven by a gradient $\nabla T \approx \Delta T/(L/2)$. 
We make use of the device's overall heat capacity $C = L\sum_{k=1}^3 A_k \rho_k c_k \doteq 3.70 \times 10^{-10}\,\text{J/K}$
to relate the change in heat energy to a corresponding change in temperature and thus to produce
an estimate of 
\begin{equation}\tau_{\text{th}} = \frac{L^2\sum_{k=1}^3 A_k\rho_k c_k}{2 \sum_{k'=1}^3 A_{k'}\kappa_{k'}} \doteq 750\,\mu\text{s}.
\end{equation}
(Values of the density $\rho_k$, specific heat $c_k$, and thermal conductivity $\kappa_k$ 
are tabulated per layer in the supplementary material.\cite{SM})
Note that $\tau_{\text{th}}$ scales as $L^2$, since the equilibration process is essentially diffusive; this implies
that, near resonance, $\omega \tau_{\text{th}} \propto L$.
Because the layers themselves come into local thermal equilibrium on a fast time scale of around
$Ct_1/Lw_1\kappa_1 \sim 5\,\text{ns}$ (where $t_1$ and $w_1$ are the thickness and width of the silicon nitride respectively),
we proceed as if 
the silicon nitride, chromium, and gold layers are
at the same temperature instantaneously at each point along the string. This is a safe assumption 
up to a frequency scale $1/(2\pi\cdot5\,\text{ns}) \approx 32~\text{MHz}$, which is well above the
oscillation frequency of any of the harmonic modes that we can measure in our device.

The simple analysis we have outlined is primarily limited by the
fact that we have supposed the temperature profile $\Delta T(x,t) = \Delta T(t)$ to have no meaningful variation along the length
of the string. Still, it reliably captures the fact that the thermal time scale is on the order of microseconds (rather than
 nanoseconds, as for the device described in Ref.~\onlinecite{Bar07}), orders of magnitude faster than the time 
 for the mechanical energy to be transferred to the environment, as dictated by the high $Q$ of the string: $Q/f_1 \approx 200$~ms.  
 
In our AC local heating scenario, the power pumped into the string is given by $P = V^2/R$, where $R$ is the string's electrical resistance and $V(t) = V_0 \cos \omega t$ is the alternating voltage applied across the string.
Since the silicon nitride is insulating ($R_1 \approx \infty$), we can estimate the resistance from the geometry
of the device using standard values $r_2$, $r_3$ for the resistivity of Cr and Au:
$R = (\sum_{k=1}^3 1/R_k)^{-1}  \approx L(A_2/r_2 + A_3/r_3)^{-1} = 49\,\Omega$.
The 176~$\Omega$ we measure experimentally is higher, mainly because it includes the resistance 
of the metallic supports.  Accordingly, the heat content of the string (ignoring the initial transient) varies in time as
\begin{equation}
U(t)=\biggl(\frac{V_0^2\tau_{\text{th}}}{2R}\biggr)\biggl[1 + \frac{\cos (2\omega t-\phi)}{(1+4\omega^2 \tau_{\text{th}}^2)\cos\phi}\biggr]
\end{equation}
with a phase lag $\phi = \tan^{-1}(2\omega \tau_{\text{th}})$ that depends only on the thermal outflow time $\tau_{\text{th}}$.
We can decompose the corresponding temperature variation into a time-invariant background $(\Delta T)_{\text{ave}} = V_0^2 \tau_{\text{th}}/2CR$ over which is superposed the weaker, alternating behaviour $(\Delta T)_{\text{alt}}$.
Hence, the analogue of Eq.~\eqref{eq:rel_freq_shift} is
\begin{equation} \label{eq:rel_freq_vs_voltage}
\begin{split}
\frac{f_n(\Delta T) - f_n(0)}{f_n(0)} &= -\frac{V_0^2\tau_{\text{th}}}{4\sigma A_1 CR}\sum_{k=1}^3 A_k E_k\alpha_k\\
&= -\frac{L}{8\sigma A_1 R}\biggl( \frac{\sum_{k=1}^3 A_kE_k\alpha_k}{\sum_{k'=1}^3 A_{k'}\kappa_{k'}}\biggr) V_0^2.
\end{split}
\end{equation}
The relative frequency shift is quadratic in $V_0$ and always negative---consistent with our observations in Fig.~\ref{fig3}(c).
In contrast to the relative frequency shift described by Eq.~\eqref{eq:rel_freq_shift},
which is insensitive to the device length, the effect described in Eq.~\eqref{eq:rel_freq_vs_voltage}
is predicted to scale as $\tau_{\text{th}}/C \sim L$ (and thus can be made arbitrary large by choosing a long enough string). 
The predicted coefficient, 6.82~V$^{-2}$, is about ten times larger than the value $0.648(9)\,\text{V}^{-2}$ obtained from a fit to the data.  Theory and experiment can be made to coincide if instead we take the thermal time to be $\tau_{\text{th}} = 72$\,$\mu$s.  The discrepancy likely stems from the over-simplified, uniform temperature profile along the length of the string.  

\begin{figure}[t]
\centerline{\includegraphics[width=3.4in]{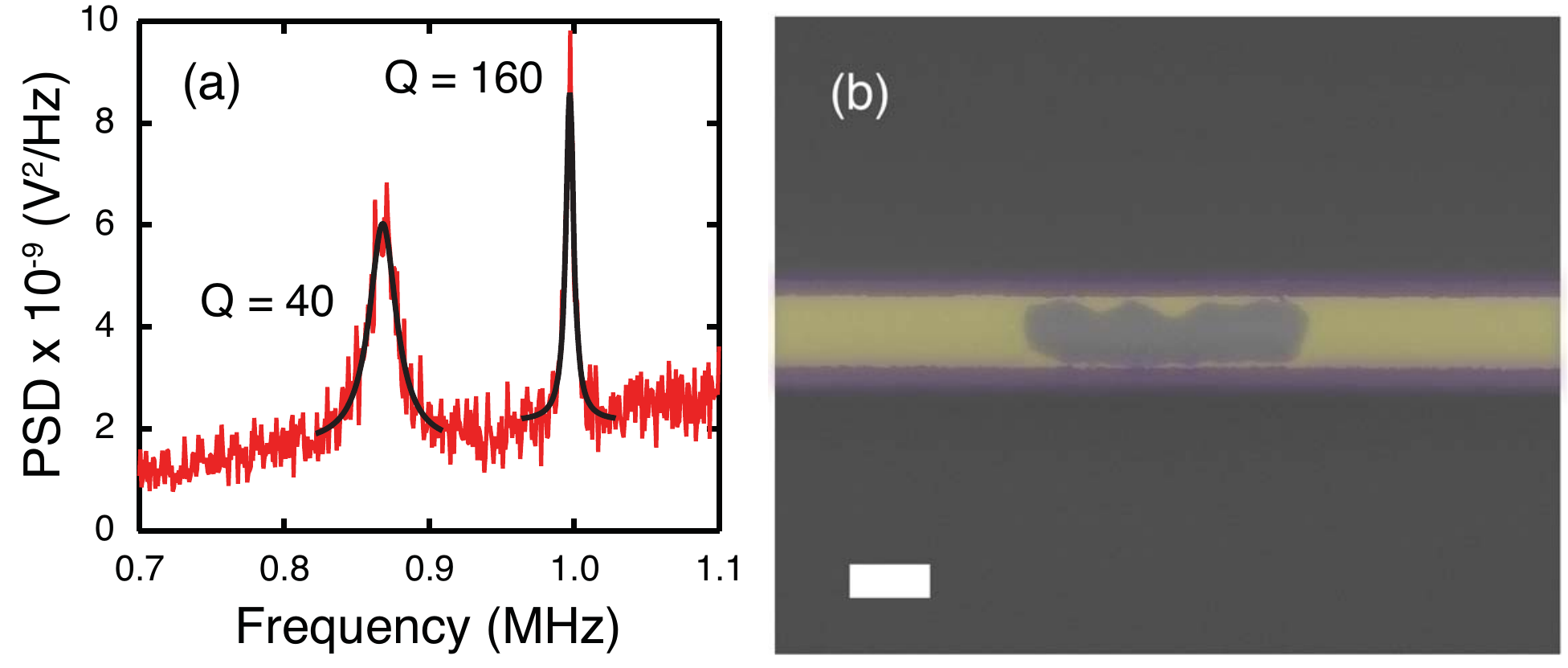}}
\caption{{\label{fig4}}  (a) Frequency response of the first out-of-plane (lower frequency) and in-plane (higher frequency) modes in air using AC thermoelastic actuation ($400$~mV drive) with corresponding PSD fits (black) to extract the $Q$.  (b)  SEM of the gold bilayer nanostring after current induced failure at $\approx 450$~mV AC drive.  Enough gold has come off the device to break the current path, resulting in insulating behavior.  Scale bar is $2~\mu$m. }
\end{figure}

In the vicinity of the fundamental mode of our device, $2\omega_1 \tau_{\text{th}} = 4\pi f_1 \tau_{\text{th}} \doteq 768 \gg 1$,
so that $\phi$ has essentially achieved its maximum value, $\pi/2$. In this limit,
$(\Delta T)_{\text{alt}} = (V_0^2/2CR\omega) \sin (2\omega t)$, and the magnitude of
the alternating component of the temperature with respect to the average
goes as $ 1/\omega\tau_{\text{th}} \ll 1$. What's astonishing is that this tiny
alternating local heating is sufficient to actuate the device and to do so very efficiently.  Specifically, we are able to use this local heating to actuate the bilayer nanostring by passing an AC current along the length of the device.  Note that this is subtly different from the bimaterial effect in cantilevers and non-uniformly coated beams, where the gold exerts a force on the cantilever that pushes it towards the substrate.\cite{Bar07,Seo08,Vil11}  In such a scenario, the force is directional, and hence the AC drive voltage must be applied at frequency $f/2$ in order to produce heating and bending at $f$. In contrast, in our system there is no preference towards or away from the substrate as the gold expands, since it forms a continuous layer over the string. Heating and bending occur twice per string oscillation, and therefore the applied drive voltage and resulting motion have the same frequency.  We are also able to actuate the device at $f/2$, although $\approx 1000$ times less efficiently---consistent with the above scenario.

In Fig.~\ref{fig3}(b), we show frequency sweeps using the reference output of the Zurich lock-in amplifier to resonantly drive the nanostring.  This integrated electronic actuation scheme is simple, compact, and robust.  With root-mean-square AC voltages as small as $100\ \mu$V we can clearly detect resonant motion, while the maximum signal-to-noise is achieved with approximately $15$ to $20$ mV.  At higher drives the nanostring becomes nonlinear.  Yet even at quite high drives, in vacuum, we see no change in the resistance of the gold layer.  Hence the local heating of the bilayer nanostring is nondestructive to the gold layer in the linear actuation regime and beyond.  In addition, we note that the resonance frequency is entirely non-hysteretic at these actuation voltages---further evidence that the AC ohmic actuation scheme is robust.  Only at voltages $>180$ mV, approximately 10 times that required for optimal signal-to-noise, is there an onset of resistance changes in the gold layer. The device continues to function with applied voltages in excess of $350$ mV, after which it rapidly becomes an open circuit.  Using the extracted value $\tau_{\text{th}} = 72~\mu$s, we can estimate that the upturn in resistance, appearing at  $V_0 \approx 180$~mV in Fig.~3(c), begins when the device has been heated an additional $\Delta T \approx 56$~K and that burn out occurs when $V_0 \approx 450$~mV and $\Delta T \approx 350$~K.  SEM of our device, shown in Fig.~\ref{fig4}(b), reveals that the gold has only been removed from a small patch on the string surface yet has interrupted the current path. 

Finally we note that the AC ohmic actuation technique works for both higher-order harmonics in vacuum (not shown) as well as for the fundamental modes in air. In the ambient pressure case, the nanostrings experience strong viscous damping, especially when the motion is perpendicular to the string's substantial width.  The $Q$ of the first out-of-plane mode is only 40 while that of the in-plane mode is 160 [see Fig.~\ref{fig4}(a)].  There is a greater cross-section for molecular collisions, and thus energy loss, in the out-of-plane direction.  The voltages required to actuate the nanostring in air ($\approx 400$~mV) result in resistance changes and eventual failure.  Improving the $Q$ in viscous environments \cite{Li07,Sun11} would decrease the required actuation voltages and prevent such failure.

In conclusion, gold bilayer nanostrings are an enhancement to the already exciting system of silicon nitride nanostrings.  They enable chemical functionalization and molecular detection, provide a scheme for local heating to study thermal response of molecules or desorb unwanted molecules, and provide simple and integrated actuation. And unlike traditional micromechanical resonators,\cite{San05} they maintain their high quality factors even with a continuous metallic coating.

This work was supported by the University of Alberta, Faculty of Science; the Canada Foundation for Innovation; the Natural Sciences and Engineering Research Council, Canada; and the Canada School of Energy and Environment.  We thank J.M. Gibbs-Davis and M.R. Freeman for helpful discussions, and thank Don Mullin, Greg Popowich and the University of Alberta NanoFab staff for technical assistance.

\setcounter{equation}{0}
\setcounter{figure}{0}
\setcounter{table}{0}
\setcounter{section}{19}
  \renewcommand{\thesection}{\Alph{section}}
\numberwithin{equation}{section}

\renewcommand{\thefigure}{S-\arabic{figure}}
\renewcommand{\thetable}{S-\Roman{table}}

\section*{Supplementary Material}

\subsection*{Device geometry}

\begin{figure}
\begin{center}
\includegraphics{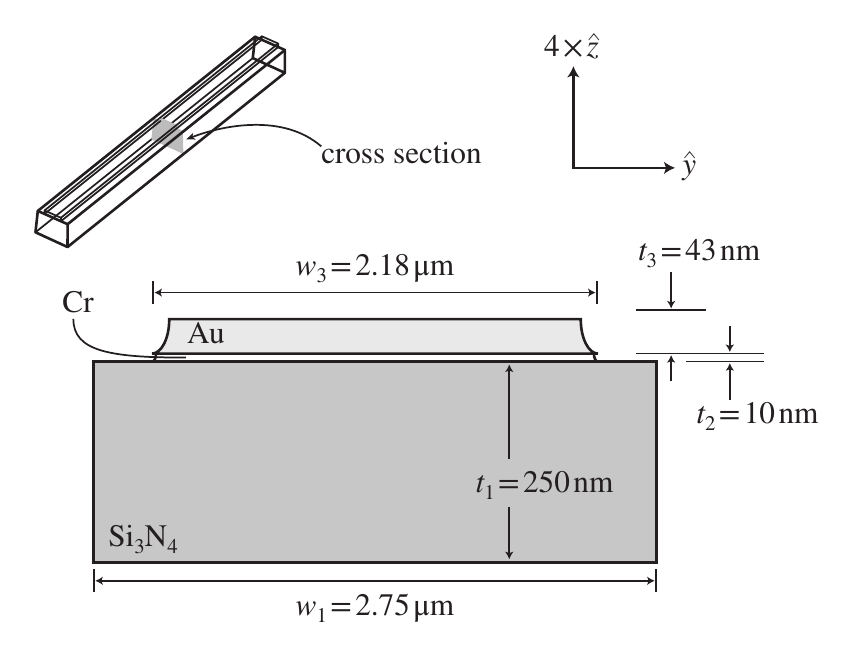}
\end{center}
\caption{\label{FIG:crossection} Bilayer nanostring viewed in cross section. Scale magnified four-fold in the vertical direction.}
\end{figure}

The device is shown in cross section in Fig.~\ref{FIG:crossection}.
Each of the silicon nitride ($k=1$), chromium ($k=2$), and gold ($k=3$) layers
is of constant thickness $t_k$ and has an area $A_k$ and a bending moment $I_k$ given by
\begin{equation}
\begin{split}
A_k &= \int_{\zeta_{k-1}}^{\zeta_k} d\zeta\,w(\zeta),\\
I_k &= \int_{\zeta_{k-1}}^{\zeta_k} d\zeta\,(\zeta-\bar{\zeta})^2 w(\zeta).
\end{split}
\end{equation}
The limits of integration are the cumulative layer thicknesses $\zeta_k = \sum_{l=1}^{k} t_l$.
The bending axis of the composite string is fixed at a height $\bar{\zeta} = 136\,\text{nm} = 0.54t_1$
(measured from the bottom surface) that minimizes the total bending constant
$\sum_{k=1}^3 E_k I_k/(1-\nu_k^2)$. Hence, the $I_k$ values themselves depend indirectly on the Young's modulus
$E_k$ and Poisson ratio $\nu_k$ of the material in each layer.

Based on our best understanding of the wet- and dry-etch steps in the fabrication process,
we propose that the layered string has a width profile of the form
\begin{equation}
w(z) = \begin{cases}
w_3 - 2\sqrt{t_3^2-(\zeta_3-z)^2} & \text{if $\zeta_2 < z < \zeta_3$}\\
w_3 - 2\sqrt{t_2^2-(\zeta_2-z)^2} & \text{if $\zeta_1 < z < \zeta_2$}\\
w_1  & \text{if $0 < z < \zeta_1$}
\end{cases}
\end{equation}
The values $w_1 = 2.75$\,$\mu$m and $w_3 = 2.18$\,$\mu$m
are estimated from a (top-down) scanning electron microscopy image of the device.
We use a pixel-contrast histogram analysis to perform edge detection.
Finally, the areas and bending moments have the following values:
\begin{equation}
\begin{split}
(A_1,A_2,A_3) &= (68.8,2.16,9.08) \times 10^{-14}\,\text{m}^2, \\
(I_1,I_2,I_3) &= (36.6,3.08,19.4) \times 10^{-28}\,\text{m}^4. 
\end{split}
\end{equation}

The device geometry and dimensions we have indicated correspond to a device of total mass 841\,pg and
heat capacity $3.70 \times 10^{-10}$\,J/K.

\subsection*{Normal Modes}

The layered beam is described by a
Lagrangian
\begin{equation}
\mathcal{L} = \int_0^L\!dx\,\biggl(\frac{M}{2L}(\partial_t z)^2 - U[z(x,t)] \biggr),
\end{equation}
where $M/L = \sum_{k=1}^3 \rho_k A_k$ is the (uniform) mass per unit length
and
\begin{multline}
U = \frac{1}{2}\sigma A_1(\partial_x z)^2 + \frac{1}{8}\sum_{k=1}^3 E_kA_k (\partial_x z)^4 \\
+ \frac{1}{2}\sum_{k=1}^3 \frac{E_k I_k}{1-\nu_k^2} (\partial_x^2 z)^2
\end{multline}
is the elastic energy density (including tension, elongation, and bending terms) 
associated with its transverse deformation.  $U$ penalizes any deviation from the equilibrium position $z(x,t) = 0$.
At this stage, we ignore
thermal expansion effects and restrict ourselves to a
model that satisfies $U[z(x,t)] = U[z(L-x,t)]$ and $U[z(x,t)] = U[-z(x,t)]$.
In other words, we assume that the system is invariant under the transformations 
$z(x) \to z(L-x)$ and $z(x) \to -z(x)$, even though the latter
symmetry can in principle be broken by the layered structure.

We follow the usual prescription
\begin{equation}
\frac{d}{dt} \frac{\delta \mathcal{L}}{\delta [\partial_t z(x,t)]} = \frac{\delta \mathcal{L}}{\delta z(x,t)}
\end{equation}
and make use of the variational identities
\begin{equation}
\begin{split}
\frac{\delta}{\delta z(x,t)} \int\!dx\,(\partial_x z)^2 &= -2\partial^2_x z\\
\frac{\delta}{\delta z(x,t)} \int\!dx\,(\partial_x z)^4 &= -12(\partial_xz)^2\partial^2_x z\\
\frac{\delta}{\delta z(x,t)} \int\!dx\,(\partial^2_x z)^2 &= 2(\partial_x z)^4
\end{split}
\end{equation}
to obtain the equation of motion 
\begin{equation} \label{EQ:equation_of_motion}
M \partial^2_t z = \tau \partial^2_x z + \frac{3}{2} S (\partial_x z)^2 \partial_x^2 z  - B\partial^4_x z.
\end{equation}
Here, $\tau = LA_1\sigma$ is the tension in the composite string, and $S = L\sum_{k=1}^3 E_k A_k$
and $B = L\sum_{k=1}^3 E_k I_k/(1-\nu_k^2)$ are its stretching and bending constants.
It is sufficient to treat $(\partial_x z)^2 \partial_x^2 z$, the term that is nonlinear in $z$,  at the mean field level. 
We do so by replacing $(\partial_x z)^2$ by its spatially averaged value
\begin{equation}
\mu = \frac{1}{L}\int_0^L \!dx\, \bigl(\partial_x z\bigr)^2.
\end{equation}

Equation~\eqref{EQ:equation_of_motion} differs from a pure wave equation by terms that are proportional
to the Young's moduli. The relevant situation here is when these are sufficiently small 
that the behavior of the system is almost perfectly string-like. 
When the system is driven in the $n$th pure string
mode $z(x,t) = a_n \sin(n\pi x/L)e^{i\omega_n t}$, it oscillates at an angular frequency
\begin{equation}
\omega_n^2 = \frac{1}{M} \biggl[ \frac{n^2\pi^2}{L^2}\biggl(\tau+\frac{3}{2}S\mu\biggr)
+ \frac{n^4 \pi^4}{L^4} B\biggr].
\end{equation}
For weak nonlinearity, 
the mean square displacement $\langle a_n^2\rangle  \sim \mu$ is small compared to $L^2 \tau/S$.
In that limit,
\begin{equation} \label{EQ:mode_frequencies}
f_n^2 = \frac{1}{4M}  \biggl( \frac{n^2}{L^2}\tau 
+ \frac{n^4 \pi^2}{L^4} B \biggr).
\end{equation}
Accordingly, high order modes exhibit a small $n^3$ correction away from
a perfectly harmonic spectrum:
\begin{equation}
f_n \approx \frac{n}{2L} \sqrt{\frac{\tau}{M}} + 
\frac{n^3\pi^2}{4L^3}\frac{B}{\sqrt{\tau M}}.
\end{equation}

\begin{table*}[t]
\caption{Material properties of the layered nanostring and substrate. Unless otherwise
specified, reference values for the bulk material are used (drawn from
\textit{CRC Handbook of Chemistry and Physics}, 92nd Edition).
}
\vspace{0.05cm}
\begin{tabular}{r.c.....} \toprule[0.75pt]
& & & &
\multicolumn{1} {c}{thermal} & 
\multicolumn{1} {c}{specific} & 
&
\multicolumn{1} {c}{thermal} \\
layer (\#)& \multicolumn{1} {c}{density} & 
\multicolumn{1} {c}{Young's modulus} & 
\multicolumn{1} {c}{Poisson ratio} & 
\multicolumn{1} {c}{expansion} & 
\multicolumn{1} {c}{heat} & 
\multicolumn{1} {c}{resistivity} &
\multicolumn{1} {c}{conductivity} \\
& \multicolumn{1} {c}{$\rho$} & 
\multicolumn{1} {c}{$E$} & 
 & 
\multicolumn{1} {c}{$\alpha$} & 
\multicolumn{1} {c}{$c$} & 
\multicolumn{1} {c}{$r$} &
\multicolumn{1} {c}{$\kappa$} \\
& \multicolumn{1} {c}{(g/cm$^3$)} & 
\multicolumn{1} {c}{(GPa)} & &
\multicolumn{1} {c}{($\times 10^{6}$~K$^{-1}$)} & 
\multicolumn{1} {c}{(J/kg$\cdot$K)} & 
\multicolumn{1} {c}{(n$\Omega$$\cdot$m)} &
\multicolumn{1} {c}{(W/m$\cdot$K)} 
\\ \midrule[0.5pt]
Au (3) & 19.30 & \phantom{2}79\phantom{$^*$} & 0.44 & 14.2  & 129.1 & 22.14 & 318\\
Cr  (2) & 7.19 & 279\phantom{$^*$} & 0.21 & 4.9 & 448.2 & 125 & 93.9 \\
Si$_3$N$_4$ (1) & 3.05\footnote{determined by multimode calibration measurements} & 265$^{\text{a}}$ & 0.20\footnote{from Ref.~\onlinecite{Khan04}} &1.27^{\text{a}} & 700 & \infty & 30 \\
Si (0) & & & & 2.59\footnote{from Ref.~\onlinecite{Oka84}} & & & 
\\ \bottomrule[0.75pt]
\end{tabular}
\end{table*}

\subsection*{Thermal Expansion}

Expansion of the layers due to heating leads to a correction that enters the elongation term of the
elastic energy density:
\begin{multline} \label{EQ:energy_elongated}
U = \frac{1}{2}A_1\sigma(\partial_x z)^2 + \frac{1}{8}\sum_{k=1}^3 E_kA_k \bigl[ (\partial_x z)^2 - 2\alpha_k \Delta T \bigr]^2\\
+ \frac{1}{2}\sum_{k=1}^3 \frac{E_k I_k}{1-\nu_k^2} (\partial_x^2 z)^2.
\end{multline}
Note that we are not attempting to model interfacial shearing and peeling stresses.  
We just assume that the layers are held together by a sufficiently large through-thickness spring constant.
Expanding Eq.~\eqref{EQ:energy_elongated} to give
\begin{multline}
U = \text{const} + \frac{1}{2L}\biggl[ \Bigl(\tau - L\sum_{k=1}^3 E_kA_k\alpha_k \Delta T\Bigr)(\partial_x z)^2\\
+ B(\partial_x^2 z)^2 + \frac{1}{4}S(\partial_x z)^4 )\biggr],
\end{multline}
we see that the effect of heating is simply to shift the tension. Applying this observation to Eq.~\eqref{EQ:mode_frequencies} gives
\begin{equation}
\begin{split}
\frac{f_n(\Delta T) - f_n(0)}{f_n(0)} &= -\frac{1}{8ML(f_n/n)^2}\sum_{k=1}^3 E_k A_k \alpha_k \Delta T\\
&\approx -\frac{1}{2\sigma A_1}\sum_{k=1}^3 E_k A_k \alpha_k \Delta T.
\end{split}
\end{equation}
The last line above is correct to the extent that $f_n/n \approx f_1$.

\subsection*{Material properties}

The properties of silicon nitride thin films are quite different from those of the bulk material.
For example, the density and coefficient of thermal expansion are
generally lower. So we have taken care to find values that are appropriate to our particular devices. 

For the purposes of calibration, we prepare a bare silicon nitride nanostring with no metallic coating.
The dimensions of this device are 215\,$\mu$m (length), 2.1\,$\mu$m (width), and 250\,nm (thickness).
The frequencies (in MHz) of the first six resonant modes are determined to be
1.19508,
2.38389,
3.58222,
4.77447,
5.98263,
and 7.19362.
In this case, Eq.~\eqref{EQ:mode_frequencies} reduces to
\begin{equation}
\begin{split}
(2Lf_n)^2 & = n^2 \frac{\sigma}{\rho}  + n^4 \pi^2 \frac{E}{\rho} \frac{(t/L)^2}{12(1-\nu^2)} \\
 & = n^2 \frac{\sigma}{\rho}  + n^4 \frac{E}{\rho} 1.16\times 10^{-6}.
\end{split}
\end{equation}
A best fit of this form to the data produces
\begin{equation}
\begin{split}
\frac{\sigma}{\rho} &= 2.622(3) \times 10^5\,\text{m$^2$/s$^2$}, \\
\frac{E}{\rho} &= 8.7(8) \times 10^7\,\text{m$^2$/s$^2$},
\end{split}
\end{equation}
and hence $E/\sigma = 330(30)$. We have no direct measurement that closes this set of relations,
but if we assume the manufacturer's quoted value of 0.8\,GPa for the tensile stress, then 
$E = 265$\,GPa and $\rho = 3.05$\,g/cm$^{3}$. These values are consistent with other measurements
in the literature.\cite{Khan04}

Furthermore, we obtain the relative frequency shift for each of modes 1, 3, and 5. The measurement is
performed over a dense grid of temperature values from 293\,K to 320\,K
and produces a linear slope
\begin{equation}
\frac{E(\alpha_0-\alpha_1)}{2\sigma} = 0.000219(1)\,\text{K}^{-1}
\end{equation}
that is consistent across all three modes.
Taking $\alpha_0 = 2.59 \times 10^{-6}$\,K$^{-1}$ as the coefficient of thermal expansion for
the silicon substrate,\cite{Oka84} we find 
$\alpha_1 = 1.27 \times 10^{-6}$\,K$^{-1}$. 
This is very close to the value $1.23 \times 10^{-6}$\,K$^{-1}$
reported in Ref.~\onlinecite{Lar11}.


\begin{thebibliography}{xxx}

\bibitem{Ver06}
S.S. Verbridge, J.M. Parpia, R.B. Reichenbach, L.M. Bellan and H.G. Craighead, \textit{J. Appl. Phys.} \textbf{99}, 124304 (2006).

\bibitem{Ver08a}
S.S. Verbridge, H.G. Craighead and J.M. Parpia, \textit{Appl. Phys. Lett.} \textbf{92}, 013112 (2008).

\bibitem{Sch11a}
S. Schmid, K.D. Jensen, K.H. Nielsen and A. Boisen, \textit{Phys. Rev. B} \textbf{84}, 165307 (2011).

\bibitem{Ver08b}
S.S. Verbridge, R. Ilic, H.G. Craighead and J.M. Parpia, \textit{Appl. Phys. Lett} \textbf{93}, 013101 (2008).

\bibitem{Unt11}
Q.P. Unterreithmeier, T. Faust and J.P. Kotthaus, \textit{Phys. Rev. Lett.} \textbf{105}, 027205 (2011).

\bibitem{Unt09}
Q.P. Unterreithmeier, E.M. Weig and J.P. Kotthaus, \textit{Nature} \textbf{458}, 1001 (2009).

\bibitem{Dav10a}
J.P. Davis, D. Vick, J.A.J. Burgess, D.C. Fortin, P. Li, V. Sauer, W.K. Hiebert and M.R. Freeman, \textit{New Journal of Physics} \textbf{12}, 093033 (2010).

\bibitem{Suh12}
A. Suhel, B.D. Hauer, T.S. Biswas, K.S.D. Beach and J.P. Davis, \textit{Appl. Phys. Lett.} \textbf{100}, 173111 (2012).

\bibitem{Ane09}
G. Anetsberger, O. Arcizet, Q.P. Unterreithmeier, R. Rivi\`ere, A. Schliesser, E.M. Weig, J.P. Kotthaus and T.J. Kippenberg, \textit{Nature Phys.} \textbf{5}, 909 (2009).

\bibitem{Fon10}
K.Y. Fong, W.H.P. Pernice, M. Li and H. X. Tang, \textit{Appl. Phys. Lett} \textbf{97}, 073112 (2010).

\bibitem{Ber97}
R. Berger, E. Delamarche, H. Peter Lang, C. Gerber, J.K. Gimzewski, E.Meyer and H.-J. G\"untherodt, \textit{Science} \textbf{276}, 2021 (1997).

\bibitem{Arl11}
J.L. Arlett,	E.B. Myers and M.L. Roukes, \textit{Nature Nano.} \textbf{6}, 203 (2011).

\bibitem{Gru11}
K. Gruber, T. Horlacher, R. Castelli, A. Mader, P.H. Seeberger and B.A. Hermann, \textit{ACS Nano} \textbf{5}, 3670 (2011).

\bibitem{Mar02}
R. Marie, H. Jensenius, J. Thaysen, C.B. Christensen and A. Boisen, \textit{Ultramicroscopy} \textbf{91}, 29 (2002).

\bibitem{God04}
M. Godin, P. J. Williams, V. Tabard-Cossa, O. Laroche, L.Y. Beaulieu, R.B. Lennox and P.H. Gr\"utter, \textit{Langmuir} \textbf{20}, 7090 (2004).

\bibitem{God10}
M. Godin, V. Tabard-Cossa, Y. Miyahara, T. Monga, P.J. Williams, L.Y. Beaulieu, R.B. Lennox and P.H. Gr\"utter, \textit{Nanotechnology} \textbf{21}, 075501 (2010).

\bibitem{San05}
R. Sandberg, K. Molhave, A. Boisen and W. Svendsen, \textit{J. Micromech. Microeng.} \textbf{15}, 2249 (2005).

\bibitem{Sos11}
G. Sosale, K. Da, L. Fr\'echette and S. Vengallatore, \textit{J. Micromech. Microeng.} \textbf{21}, 105010 (2011).

\bibitem{Lab12}
A. Labuda, J.R. Bates and P.H. Gr\"utter, \textit{Nanotechnology} \textbf{23}, 025503 (2012).

\bibitem{Gim94} J.K. Gimzewski, Ch. Gerber, E. Meyer and R.R. Schlittler, \textit{Chemical Physics Letters} \textbf{217}, 589 (1994).

\bibitem{Bar94} J.R. Barnes, R.J. Stephenson, C.N. Woodburn, S.J. O'Shea, M.E. Welland, T. Rayment, J.K. Gimzewski and Ch. Gerber, \textit{Rev. Sci. Instrum.} \textbf{65}, 3793 (1994).

\bibitem{Var97} J. Varesi, J. Lai, T. Perazzo, Z. Shi and A. Majumdara, \textit{Appl. Phys. Lett} \textbf{71}, 306 (1997).

\bibitem{Lar11}
T. Larsen, S. Schmid, L. Gr\"onberg, A.O. Niskanen, J. Hassel, S. Dohn and A. Boisen, \textit{Appl. Phys. Lett} \textbf{98}, 121901 (2011).

\bibitem{Cha12}
J. Chaste, A. Eichler, J. Moser, G. Ceballos, R. Rurali and A. Bachtold, \textit{Nature Nano.} \textbf{7}, 301 (2012). 

\bibitem{Ber98}
R. Berger, H.P. Lang, Ch. Gerber, J.K. Gimzewski, J.H. Fabian, L. Scandella, E. Meyer and H.-J. G\"untherodt, \textit{Chemical Physics Letters} \textbf{294}, 363 (1998).

\bibitem{Zha11}
J. Zhao, X. Yin, J. Shi, X. Zhao and J.S. Gutmann, \textit{J. Phys. Chem. C} \textbf{115}, 22347 (2011).

\bibitem{Ier11}
E. Iervolino, A.W. van Herwaarden, W. van der Vlist and P.M. Sarro, \textit{J. Microelectromechanical Systems} \textbf{20}, 1277 (2011).

\bibitem{Liu12}
T. Liu, S. Pihan, M. Roth, M. Retsch, U. Jonas, J.S. Gutmann, K. Koynov, H.-J. Butt and R. Berger, \textit{Macromolecules} \textbf{45}, 862 (2012).

\bibitem{Yi08}
D. Yi, A. Greve, J.H. Hales, L.R. Senesac, Z.J. Davis, D.M. Nicholson, A. Boisen and T. Thundat, \textit{Appl. Phys. Lett.} \textbf{93}, 154102 (2008).

\bibitem{Bar07}
I. Bargatin, I. Kozinsky and M.L. Roukes, \textit{Appl. Phys. Lett.} \textbf{90}, 093116 (2007).

\bibitem{Seo08}
J. H. Seo and O.  Brand, \textit{J. Microelectromech. Syst.} \textbf{17}, 483 (2008).

\bibitem{Vil11}
L.G. Villanueva, R.B. Karabalin, M.H. Matheny, E. Kenig, M.C. Cross and M.L. Roukes, \textit{Nano. Lett.} \textbf{11}, 5054 (2011).

\bibitem{Dav10b}
J.P. Davis, D. Vick, D.C. Fortin, J.A.J. Burgess, W.K. Hiebert and M.R. Freeman, \textit{Appl. Phys. Lett.} \textbf{96}, 072513 (2010).

\bibitem{Wil11}
I. Wilson-Rae, R.A. Barton, S.S. Verbridge, D.R. Southworth, B. Ilic, H.G. Craighead and J.M. Parpia, \textit{Phys. Rev. Lett.} \textbf{106}, 047205 (2011).

\bibitem{Moh02} 
P. Mohanty, D.A. Harrington, K.L. Ekinci, Y.T. Yang, M.J. Murphy and M.L. Roukes, \textit{Phys.\ Rev.\ B} \textbf{66}, 085416 (2002).

\bibitem{Doh07} 
S. Dohn, W. Svendsen, A. Boisen, and O. Hansen, \textit{Rev. Sci. Instrum.} \textbf{78}, 103303 (2007).

\bibitem{Sch10}
S. Schmid, S. Dohn and A. Boisen, \textit{Sensors} \textbf{10}, 8092 (2010).

\bibitem{SM}
See supplementary material at [URL will be inserted by AIP] for details of the device's geometry and
material composition.

\bibitem{Li07}
M. Li, H.X. Tang and M.L. Roukes, \textit{Nature Nano.} \textbf{2}, 114 (2007).

\bibitem{Sun11}
X. Sun, K.Y. Fong, C. Xiong, W.H.P. Pernice and H.X. Tang, \textit{Optics Express} \textbf{19}, 22316 (2011).

\bibitem{Khan04} A.\ Khan, J.\ Philip and P.\ Hess, \textit{J.\ Appl.\ Phys.} {\bf 95}, 1667 (2004).

\bibitem{Oka84} Y.\ Okada and Y.\ Tokumaru, \textit{J.\ Appl.\ Phys.} {\bf 56}, 314 (1984).

\bibitem{Lar11} T. Larsen, S. Schmid, L. Gr\"onberg, A.O. Niskanen, J. Hassel, S. Dohn 
and A. Boisen, \textit{Appl. Phys. Lett} \textbf{98}, 121901 (2011).

\end{thebibliography}
\end{document}